\documentclass[10pt, preprint]{emulateapj}

\usepackage{color}
\usepackage{amsmath}	

\usepackage{style_wei_symbols}
\usepackage{style_wei}

\usepackage{natbib}
\begin{document}

\title{Episodic X-ray Emission Accompanying the Activation of an Eruptive Prominence: 
Evidence of Episodic Magnetic Reconnection}




\author{Wei Liu\altaffilmark{1}$^,$\altaffilmark{2}$^,$\altaffilmark{3},
 Tong-Jiang Wang\altaffilmark{1}$^,$\altaffilmark{4},
 Brian R.~Dennis\altaffilmark{1}, and Gordon D.~Holman\altaffilmark{1}}

\altaffiltext{1}{Solar Physics Laboratory, Code 671, Heliophysics Science Division, 
 NASA Goddard Space Flight Center, Greenbelt, MD 20771}	
\altaffiltext{2}{NASA Postdoctoral Program Fellow}
\altaffiltext{3}{Current address: Lockheed Martin Solar \& Astrophysics Lab, Dept.~ADBS, 
Bldg.~252, 3251 Hanover St., Palo Alto, CA 94304; weiliu[at]sun(.)stanford(.)edu}
\altaffiltext{4}{Department of Physics, Catholic University of America, Washington, DC 20064}

\shorttitle{Episodic X-ray Emission Accompanying an Eruptive Prominence}
\shortauthors{Liu et al.}

\begin{abstract}	

We present an X-ray imaging and spectroscopic study of a partially occulted (N16W93) C7.7 flare on 2003 April 24
observed by \hsi that accompanied a prominence eruption observed by \traceA. 
 (1) The activation and rise of the prominence occurs during the preheating phase of the flare. 
The initial X-ray emission appears as a single coronal source	
at one leg of the prominence and it then splits into a double source.
Such a source splitting happens three times, 
each coinciding with an increased X-ray flux and plasma temperature, 
suggestive of fast reconnection in a localized current sheet and an enhanced energy release rate.
In the late stage of this phase, the prominence displays a helical structure.
These observations are consistent with the tether-cutting and/or kink instability model for triggering solar eruptions.
 (2) The eruption of the prominence takes place during the flare impulsive phase.
Since then, there appear signatures predicted by		
the classical CSHKP model of two-ribbon flares occurring in a vertical current sheet trailing an eruption. 	
These signatures include an EUV cusp and current-sheet-like feature (or ridge) above it.
There is also X-ray emission along the EUV ridge both below and above the cusp, which in both regions
appears closer to the cusp at higher energies in the {\it thermal} regime ($\lesssim$20~keV). 
This trend is reversed in the {\it nonthermal} regime. 
 (3) Spectral analysis indicates thermal X-rays from all sources throughout the flare, while during
the impulsive phase there is additional nonthermal emission which primarily comes from 
the coronal source below the cusp. This source also has a lower temperature ($T=20 \pm 1$ vs.~$25 \pm 1 \MK$), 
a higher emission measure (${\rm EM}=[3.3 \pm 0.4]$ vs.~$[1.2 \pm 0.4] \E{47} \pcmc $),
and a much harder nonthermal spectrum (electron power-law index 
$\delta=5.4 \pm 0.4$ vs.~$8 \pm 1$) than the upper sources.

\end{abstract}


\section{Introduction}
\label{sect_intro}

Eruptions of solar prominences (or filaments) 
are frequently associated with and physically related to coronal mass ejections (CMEs) and flares
\citep{Tandberg-HanssenE.prominenceBook.1995nsp..book.....T}. 
Investigation of prominence eruptions can provide critical clues not only to 
prominence activity in its own right, but also to the physics of CMEs and flares.
Several proposed mechanisms of prominence eruptions fall into two categories:
(1) an ideal magnetohydrodynamic (MHD) process, such as the kink instability of a flux rope 
\citep[e.g.,][]{FanGibson.emrgKink.2003ApJ...589L.105F, Torok.Kliem.kink.2005ApJ...630L..97T,
WilliamsDR.kink-promin.2005ApJ...628L.163W}, that does not require magnetic reconnection,
and (2) a nonideal MHD process in which magnetic reconnection plays an important role. 
The latter category includes the flux cancellation \citep{vanBallegooijenMartens.promin.1989ApJ...343..971V}
or tether-cutting \citep{MooreRoumeliotis.origTether-cutting.1992LNP...399...69M} model
in which reconnection occurs {\it below} the filament, and the breakout model 
\citep{AntiochosS.breakout.1998ApJ...502L.181A, LowZhang.2classCME.2002ApJ...564L..53L}
in which reconnection takes place {\it above} the filament and removes the confinement
from the overlying field lines. On occasions when a filament eruption is accompanied by a flare,
the flare X-ray emission provides important timing and spatial information about magnetic reconnection. 
Such information, especially during the early stage of the filament activation
or the flare precursor \citep{vanHoven.Hurford.precursor.1984AdSpR...4...95V,
Harrison.etal.precursor.1985SoPh...97..387H, Harrison.precursor.1986A&A...162..283H}, is very useful in distinguishing
between the above models. 	

With soft and hard X-ray data from the \yohkoh satellite, observational evidence
has been found in support of the tether-cutting model \citep{MooreR.tether-cutting.2001ApJ...552..833M},
the breakout model \citep{WangTJ.baldpatch.2002ApJ...572..580W,
SterlingMoore.external-recon.2004ApJ...602.1024S}, or both models
\citep{SterlingMoore.exter-internal-recon.2004ApJ...613.1221S}.
Using data 	
from the {\it Reuven Ramaty High Energy Solar Spectroscopic Imager} (\hsiA), 
\citet{JiH.FailedEruption.2003ApJ...595L.135J} found simultaneous X-ray emission above and below
a filament prior to its failed eruption during which the filament became kinked.
They interpreted the higher altitude X-ray source as evidence of the breakout model. 
In the same event, \citet{AlexanderLiuGilbert.failed.2006ApJ...653..719A} found additional X-rays 
emitted from the crossing of the two legs of the kinked filament.
\citet{ChiforC.filamentErupt.2006A&A...458..965C, ChiforC.XRprecursor.2007A&A...472..967C}
examined \hsi X-ray sources during the early stages of filament eruptions	
that favored the tether-cutting model. 
At the same time, \citet{Sui.enigma.2006ApJ...646..605S} reported possible signatures of the breakout model in
an event involving multiple-loop interactions.
In brief, evidence from these observations does not seem to converge on any one model,
and which mechanism is predominant remains an open question.

In previous studies of filament eruptions accompanied by flares (except \citealt{LiuAlexander.kink.HXR.2009ApJ}), 
due to instrumental or observational limitations on imaging spectroscopy,
very little attention was paid to the combined morphology and spectra of X-ray sources, which can provide
crucial constraints for theoretical models. 
Instrumental limitations include the energy resolution (e.g., only four broad bands 
from 14 to 93~keV for the \yohkoh Hard X-ray Telescope vs.~$\sim$1~keV FWHM for \hsiA)
and dynamic range (e.g., $\lesssim$50:1 for \hsiA) of X-ray imagers.
Unfavorable observational conditions include coronal sources being too weak
to allow for imaging spectroscopy \citep[e.g.,][]{Sui.enigma.2006ApJ...646..605S},
especially in the presence of bright footpoint sources because of the limited dynamic range.
In an attempt to fill this gap, we report here
a detailed investigation of both morphology and spectra of	
the X-ray sources in a partially {\it occulted} flare
observed by \hsiA, which accompanied a prominence eruption observed by the 
{\it Transition Region and Coronal Explorer} (\traceA).		
The occultation of the footpoint X-rays enabled us to detect the relatively faint 
emission high in the corona.
The location near the limb helped minimize projection effects, 
which are a concern for events occurring on the solar disk. 

One of our significant findings is that there are 	
episodic morphological changes from a single coronal X-ray source to a double coronal source 	
during 	
the filament activation. Each occurrence of the double source structure coincides
with an increased X-ray flux and plasma temperature, 
thus suggestive of an increased reconnection or energy release rate.
Most of the time, the X-ray sources are located below the prominence apex, more in favor of the 
tether-cutting model than the breakout model.	
We present the observations in \S~\ref{sect_obs}, 
and conclude this paper with a summary and discussion in \S~\ref{sect_conclude}.


\section{Observations and Data Analysis}
\label{sect_obs}

The event under study occurred at $\sim$15:30~UT on 2003 April 24 in AR~10339, which was very active, 
producing 10 C-class flares within two days from April 23 to 24.
At the time of this event, AR~10339 had just rotated over the west limb. 
Its center is estimated to be at N16W93, 
whose occulted position on the sky plane is $3\arcsec \pm 3\arcsec$		
below the limb, according to the solar rotation and the AR location in a magnetogram taken	
one day earlier by the {\it Solar and Heliospheric Observatory} (\sohoA) Michelson Doppler Imager (MDI).
This event involved a prominence eruption accompanied by a C7.7 flare whose X-ray emission from the footpoints
is occulted by the limb. There was no CME in a $\pm2$ hour window according to the \soho LASCO catalog.
This event was well observed by \hsi and \traceA. 
Complimentary data with lower cadence and/or spatial resolution were recorded
by the \soho Extreme Ultraviolet (EUV) Imaging Telescope (EIT)
in EUV, by the Improved Solar Observing Optical Network (ISOON) in \HaA, 
and by the {\it Geostationary Operational Environment Satellite} (\goesA) Solar X-ray Imager (SXI) in soft X-rays.

%
 \begin{figure}[thbp]      
 \epsscale{1.1}	
 \plotone{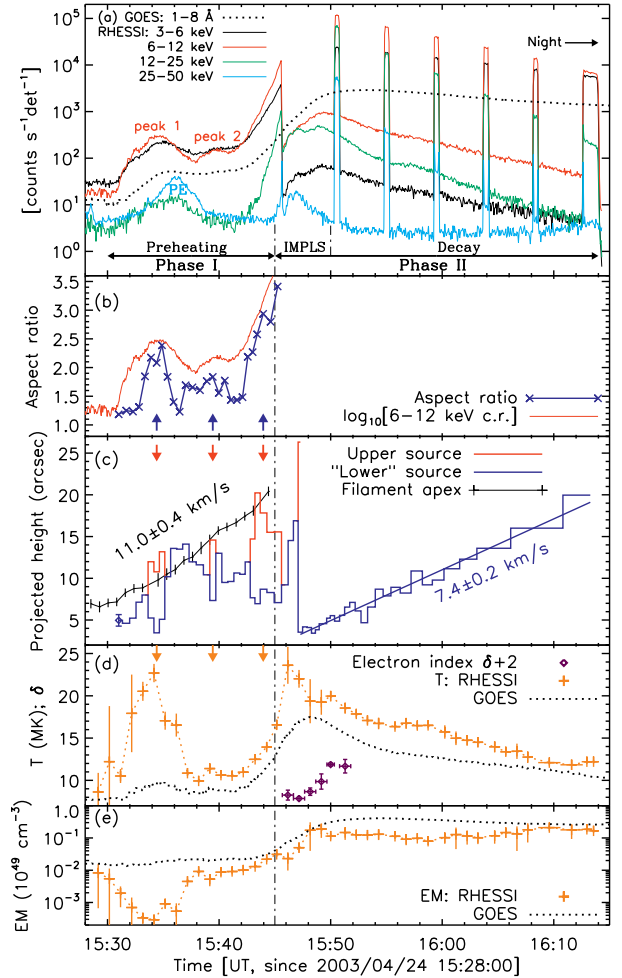}
 \caption[]{
 Temporal evolution of various quantities.
   ({\it a}) \hsi count rates	
 averaged over the front segments of 
 detectors 1, 3, 4, 6, 8, and 9, and \goes low channel flux $f_{\rm GOES}$
 in units of [W $\m^{-2}$], which is rescaled by the formula $(f_{\rm GOES}- 4.4\E{-7})\E{8}$.
 The vertical dot-dashed line at 15:45~UT divides the flare into two phases.
 ``PE" labels the 25--50 keV hump caused by a particle event.
 The downward step 	
 near 16:14~UT marks the beginning of spacecraft night.
   ({\it b}) Aspect ratio or degree of elongation
 of the X-ray emission as defined in \S\ref{subsect_phsI}.
 The data points are averaged over the 
 3--6 and 6--10~keV bands. The red curve is a portion of the logarithmic 6--12~keV count rate 
 shown in ({\it a}).
   ({\it c}) Projected height $h$ of the lower ({\it blue}) and upper ({\it red}) source centroids measured 
 along the fiducial direction defined in Fig.~\ref{mapsII.eps}{\it c}. The straight line indicates a linear fit that
 yields a velocity of $7.4\pm0.2 \km \ps$. The three arrows here and in ({\it b}) mark three
 episodes of the occurrence of a double source and the increase of source elongation.
 The dark curve is the projected height of the filament apex seen in EUV, 	
 with the vertical bars representing $\pm$1$\sigma$ uncertainties (see \S\ref{subsect_phsI}).
   ({\it d}) Plasma temperature ({\it plus signs}) and nonthermal electron spectral index 
 ({\it diamonds}, shifted up by 2)	
 inferred from fits to spatially integrated spectra. The dotted line
 is the temperature inferred from \goes data.
   ({\it e}) Same as ({\it d}) but for emission measure.	
 } \label{vs-time.eps}
 \end{figure}

Figure~\ref{vs-time.eps}{\it a} shows \hsi and \goes X-ray light curves for this event.
The interval of 15:45--15:50~UT with significant signal in the 25--50~keV channel above the background%
  \footnote{The hump in the 25--50~keV curve peaking at 15:36~UT indicates a particle event (PE)
  from Earth's radiation belt, which usually has a negligible contribution at low energies.
  When fitting spectra before 15:42~UT, we thus restricted ourselves to $E \leq 13 \keV$,
  above which the spectra flatten due to particle contamination.
  Particles have no effect on images other than adding noise, since they are not modulated by the grids.
  The method using the count rate ratio between the front and rear segments to remove particle counts
  \citep{LiuW_FPAsym_2009ApJ...693..847L}	
  is generally good for $\sim$20--150~keV or large (M--X class) flares, and is not attempted here.
  }
is referred to as the impulsive phase (labeled ``IMPLS"). The preceding interval (15:30--15:45~UT) is
the preheating phase \citep[e.g.,][]{Harrison.precursor.1986A&A...162..283H}, 
which we call Phase~I and, according to the 6--12~keV count rate,
includes two peaks and the rapid rise leading to the impulsive phase. 
As we will see later, this event exhibits various distinctions
before and after the onset of the impulsive phase, and thus we call the combination of the
impulsive and decay phases Phase~II (15:45--16:14~UT).

We reconstructed \hsi images in wide energy bands, 3--6, 6--10, and 10--15~keV,	
throughout the flare with the integration time (of integer numbers of the spacecraft spin period, $\sim$4~s) 
ranging from 24~s to 158~s, depending on the count rates. We used the front segments of
detectors~3--8 giving a FWHM resolution of $9\farcs 8$.
Shortly into the impulsive phase, \hsiA's thin attenuator moved in at 15:45:36~UT, raising the
lower end of the measurable energy range from 3 to 6~keV, except for those short intervals
in the open attenuator state (appearing as narrow steps in Fig.~\ref{vs-time.eps}{\it a}).
We primarily used the computationally expensive PIXON \citep{MetcalfT1996ApJ...466..585M} 
and faster CLEAN \citep{HurfordG2002SoPh..210...61H} algorithms for times before 
and after 15:45:36~UT, respectively.
We cross-checked both algorithms in all cases.

We analyzed spatially integrated \hsi spectra throughout the flare following the procedures
detailed in \citet[][see their Appendix~A1]{LiuW_2LT.2008ApJ...676..704L}. One important step
was to fit the spectra of individual detectors (excluding detectors 2, 5, and 7) separately,%
 \footnote{Treating detectors separately allows us to take advantage of statistically 
 independent measurements of the same incident photon spectrum made by \hsiA's 
 nine nominally identical detectors. This also helps avoid the energy smearing 
 inherent in the default procedure of averaging counts from different detectors
 that have slightly ($\sim$10\%) different energy bin edges and sensitivities.
 Detectors 2, 5, and 7 were not used for this analysis because of their higher
 energy thresholds and/or degraded energy resolution \citep{SmithD2002SoPh..210...33S}.
 }
and then average the results to obtain the best-fit parameters and use the standard deviations
of the results as uncertainties. We also applied the pileup correction ({\it pileup\_mod})
recently developed by R.~Schwartz in the {\it SolarSoft} ({\it SSW}) package.
In the attenuator open state before 15:45:36~UT, the spectra were fitted with two isothermal components,
with the lower-temperature component representing the emission from a preceding flare elsewhere (N17W39) 
on the Sun whose hot plasma still undergoes cooling. When the thin attenuator is in (after 15:45:36~UT),
we fitted the spectra with an isothermal plus nonthermal power-law mean electron flux 
\citep[][equivalent to the thin-target function in OSPEX]{Brown.meanElecflux.2003ApJ...595L.115B} model.


\trace had generally good coverage at 195~\AA\ during this event except two $\sim$7~minute data gaps before and after 15:52~UT.
\trace data were processed with the standard {\it SSW} routines and radiation spikes were removed.
To correct for the pointing offset, all \trace images were shifted by $\Delta x=-7\farcs 0$ 
and $\Delta y=-2\farcs 1$ in the solar east-west ($x$) and south-north ($y$) 
directions, respectively, according to the cross-correlation between two neighboring \trace (16:00:59~UT)
and EIT (16:00:02~UT) 195~\AA\ images \citep{GallagherP2002SoPh..210..341G}.
This coalignment is good to $\pm 1\arcsec$ in both directions. 


In the following two subsections we scrutinize the spatial and spectral evolution of the X-ray sources
and the associated prominence activity during Phases~I and II individually.

\subsection{Phase I (15:30--15:45~UT): Flare Preheating and Prominence Activation}
\label{subsect_phsI}

%
 \begin{figure*}[thbp]      
 \epsscale{0.8}
 \plotone{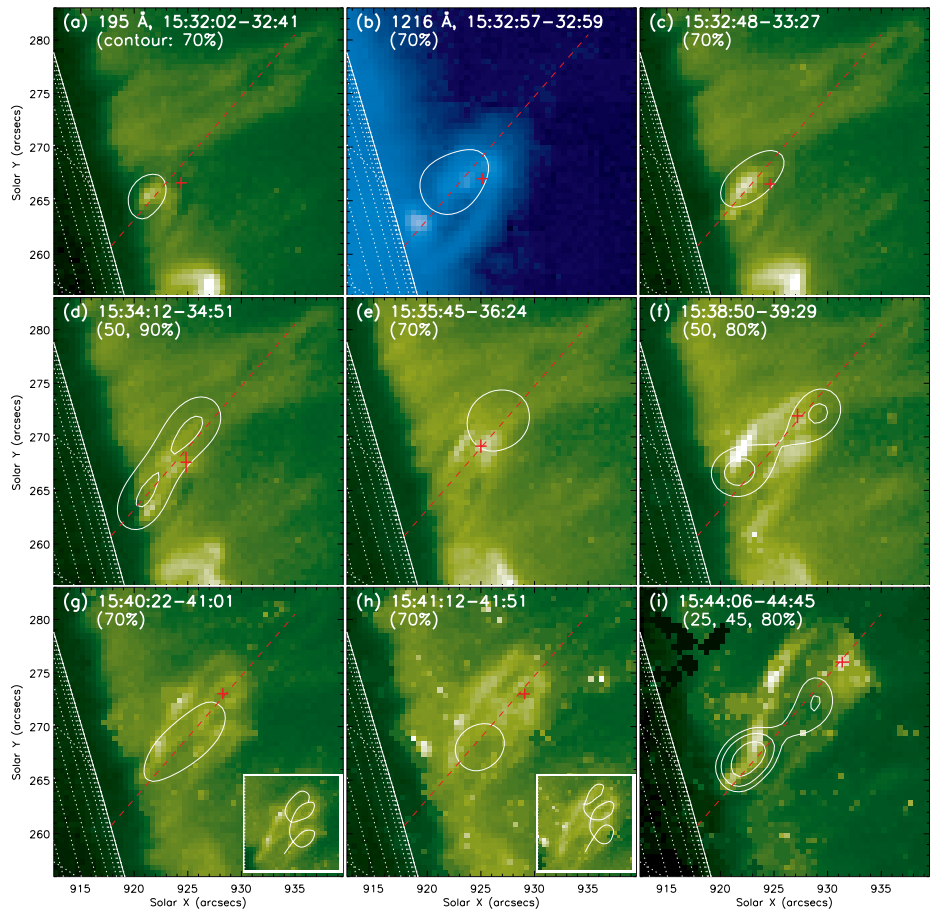}
 \caption[]{
 Selected \trace 1216 \AA\ (panel {\it b}) and 195 \AA\ (other panels) maps 
 labeled by their exposure intervals 	
 during Phase~I. Superimposed are \hsi PIXON image contours at 3--6~keV integrated 
 in 40~s intervals, whose central times coincide with those of the corresponding \trace maps. 
 Contour levels at percentages of the image maximum are given in parentheses.
 The filament apex is marked by the plus sign whose horizontal and vertical extents represent the
 positional uncertainty in each direction. 
 The dashed line, as defined in Fig.~\ref{mapsII.eps}{\it c}, 
 indicates the fiducial direction for tracking source motions (see \S\ref{subsect_phsI}).
 Panel {\it b} shows the prominence as loop-shaped emission at 1216~\AA,
 where the bright spot below the northern leg is, most likely,
 an UV continuum brightening on the ghost limb that appears $\sim$$2\arcsec$ above the true limb
 \citep[][their Fig.~5]{HandyB.Lyman-alpha.1999SoPh..190..351H}.
 The inserts in ({\it g}) and ({\it h}) offer a zoomed view of the helical structure of the prominence.
 (GIF movies of this figure can be found at http://sun.stanford.edu/$\sim$weiliu/movies/filamt-20030424)
 } \label{traceMosaic.eps}
 \end{figure*}
Phase~I is the preheating period of the flare during which the prominence undergoes activation.
Figure~\ref{traceMosaic.eps} shows the evolution of the prominence during this phase as seen by \trace in EUV
and of the X-ray emission seen by \hsiA.  We find that simultaneous and cospatial X-ray and EUV brightenings 
appear at the northern leg of the prominence as early as 15:31:16~UT (e.g., Fig.~\ref{traceMosaic.eps}{\it a}).
Both emissions rapidly become elongated along the leg (Fig.~\ref{traceMosaic.eps}{\it c}),	
and meanwhile the prominence apex, marked by the plus sign, gradually moves upward. At 15:34:12 (Fig.~\ref{traceMosaic.eps}{\it d})
the X-ray emission splits into a double source, which we call the {\it lower} and {\it upper coronal sources},
the latter of which 
extends above the prominence apex. The X-ray emission then reverts 
back to a single source (Fig.~\ref{traceMosaic.eps}{\it e}), located above the apex of the prominence, 
while the EUV emission spreads from the northern
leg to other parts of the prominence. Through the rest of the preheating phase, the morphological
transition of X-ray emission from a single source to a double source happens 
twice more	
(Figs.~\ref{traceMosaic.eps}{\it f} and \ref{traceMosaic.eps}{\it i}).
During its evolution, the prominence gradually unveils a helical structure in its top portion,
which is best seen as three bright coils (Figs.~\ref{traceMosaic.eps}{\it g} 
and \ref{traceMosaic.eps}{\it h}). (This is similar to that reported by
\citet{LiuKurokawa.filament.2004PASJ...56..497L} and \citet{LiJing.20715X3.2005ApJ...620.1092L},
except that the helical feature in their case appears during the flare impulsive phase 
and is located significantly below the erupting filament.)
The prominence was last seen at 15:44:06--15:44:45~UT 
(Fig.~\ref{traceMosaic.eps}{\it i}) just before the \trace data gap.

We tracked emission centroids to investigate X-ray source motions. Before and after the
thin attenuator came in at 15:45:36~UT, images at 3--6 and 6--10~keV were	
used, respectively. We utilized contours at 50\% of the image maximum to locate the centroids,
except that on occasions of a double source before 15:45:36~UT we employed
two independent contours at within 5\% of the minimum between the two sources.
The resulting centroid locations are plotted in Figure~\ref{mapsII.eps}{\it c} 
and fitted with a straight line ({\it dashed}).		
This line, at $33\degree$ from the local vertical on the limb, defines the {\it fiducial direction} or {\it main direction of motion} 
of all features discussed in this paper.
Here we define ``projected height" $h$ of any feature as the projection in the fiducial direction
of the distance from ``S", the crossing of the fiducial line with the limb.
The projected heights of the centroids of the lower and upper coronal sources are shown in 
Figure~\ref{vs-time.eps}{\it c}, and the standard deviation of the distance (perpendicular) 
to the fiducial line is used as the uncertainty shown on the first data point. 
Note that, at times of a single source, the centroid is assigned to the ``lower" source in Figure~\ref{vs-time.eps}.

There are three episodes of the splitting of a single source into a double source
during Phase~I, as marked by the vertical arrows (Fig.~\ref{vs-time.eps}{\it c}, see also Fig.~\ref{traceMosaic.eps}).
This splitting can also be represented by an increase in the {\it aspect ratio} 
of the entire X-ray emission, 
defined as the maximum of the ratio of the standard deviation (or second moment) of the source region
along any two orthogonal directions.
The aspect ratio averaged between the results obtained from 3--6 and 6--10~keV images before 15:45:36~UT is shown
in Figure~\ref{vs-time.eps}{\it b} 
(while this analysis was not attempted after 15:45:36~UT due to complex source morphology as we will see later).
We find a positive temporal correlation between the aspect ratio and X-ray flux.
Specifically, the three episodes (marked by three arrows) of large values of the aspect ratio
coincide with the first two X-ray peaks and the rapid rise leading to the impulsive phase.
Spectral analysis indicates high temperatures at these times (Fig.~\ref{vs-time.eps}{\it d}), 
and particularly the temperature at the first two X-ray peaks reaches its local maxima 
of $23 \pm 1 \MK$ (15:33:40--15:34:40~UT) 	
and $11.4 \pm 0.6 \MK$ (15:38:40--15:39:40~UT).
These observations suggest higher magnetic reconnection rates or energy release rates occurring at times of
the morphological transition from a single source to a double source.
A more extensive discussion on this is offered in \S~\ref{sect_conclude}.

To quantify the motion of the prominence, we tracked the center of the absorption feature at
the prominence apex or its best estimate based on morphological interpolation between the two legs. 
In the late stage (since 15:37~UT) when the absorption
is obscure but the helical structure is more evident,	
 we followed the center of the central coil.
We repeated this with four independent measurements and used their average as the final result.
The square root of the quadrature sum of the standard deviation of the measurements and the \trace
$0\farcs 5$ FWHM is used as the uncertainty in each of the $x$ and $y$ directions.
The resulting apex locations and their uncertainties are shown as plus signs in Figure~\ref{traceMosaic.eps}.
They were then projected onto the fiducial direction defined above 
(shown also in Fig.~\ref{traceMosaic.eps}) to find
their projected heights above the limb,	
whose history is shown in Figure~\ref{vs-time.eps}{\it c}.
As can be seen, the prominence shows no significant motion until the onset (15:30~UT) of the flare preheating phase. 
A linear fit to the projected filament height between 15:30 and 15:44~UT indicates a velocity of $11.0 \pm 0.4 \km \ps$,
similar to those ($<$$10\km \ps$) observed in the filament slow-rise phases of other events
\citep[e.g.,][]{MooreSterling.2006GMS...165...43M, ChiforC.XRprecursor.2007A&A...472..967C}.

Finally, we compare the positions of the X-ray sources relative to the prominence
in greater detail as shown in Figure~\ref{vs-time.eps}{\it c}. The X-rays are initially
emitted below the prominence apex. The projected height of the apex then resides between the two X-ray sources
during the first double-source episode (15:33:38--15:35:07~UT), followed by a period
(15:35:40--15:37:17~UT) when the apex is located below the single source (by $2\farcs 5 \pm 0\farcs 9$ at 15:36~UT).
During the rest of Phase~I when the prominence has risen well above
the limb ($h \geq 13 \farcs 6 \pm 0\farcs 7$), the X-ray sources, in either a single or a double
source morphology, are primarily below the prominence apex.
We discuss the implications of these observations in \S~\ref{sect_conclude}.

\subsection{Phase II (15:45--16:14~UT): Flare Impulsive and Decay Phases and Prominence Eruption}
\label{subsect_phsII}

%
 \begin{figure*}[thbp]      
 \epsscale{1}
 \plotone{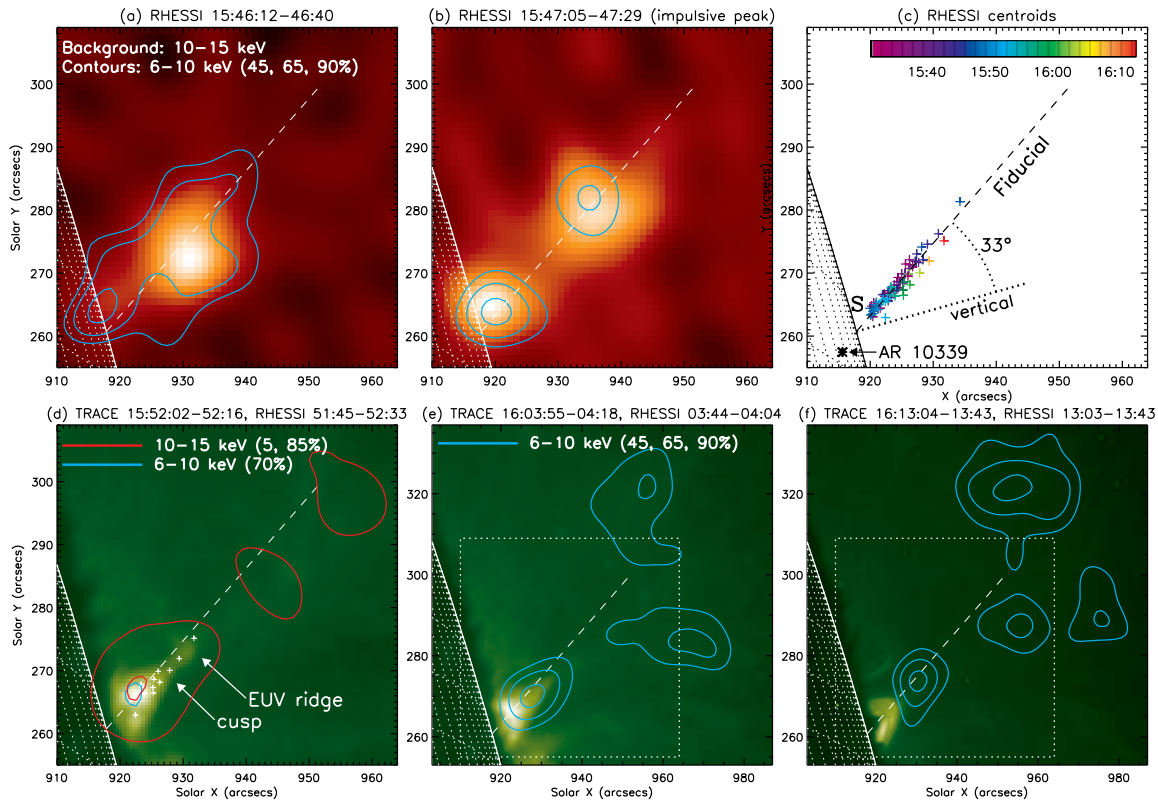}
 \caption[]{
 Evolution of X-ray and EUV sources during Phase~II. 
   ({\it a}) and ({\it b}) \hsi images at 6--10 ({\it contours}) and 10--15~keV ({\it background})
 during the time of a \trace data gap.
   ({\it c}) \hsi source centroids at 3--6~keV during 15:30:40--15:45:36~UT and 
 at 6--10~keV during 15:45:40--16:13:53~UT. Colors indicate time. The dashed line 
 (replotted in every panel) is a linear fit to the data,
 representing the fiducial direction of motion. ``S" marks its crossing with the limb,
 as the origin of projected height $h$.
 The asterisk indicates the estimated location of the center of AR~10339 behind the limb.
   ({\it d})--({\it f}) \trace 195 \AA\ maps overlaid with simultaneous \hsi image contours
 ({\it red}: 10--15~keV in panel~{\it d} only, {\it cyan}: 6--10~keV). The white plus signs in ({\it d}) 
 are a portion of the centroids shown in ({\it c}) at times after 15:52~UT.	
 The white dotted box in ({\it e}) and ({\it f}) depicts the smaller field of view (FOV) of ({\it a})--({\it d}).
 \hsi images were made with the CLEAN algorithm, except for panel {\it d} where PIXON was used.
 } \label{mapsII.eps}
 \end{figure*}
\begin{figure}[thbp]      
 \epsscale{1.1}	
 \plotone{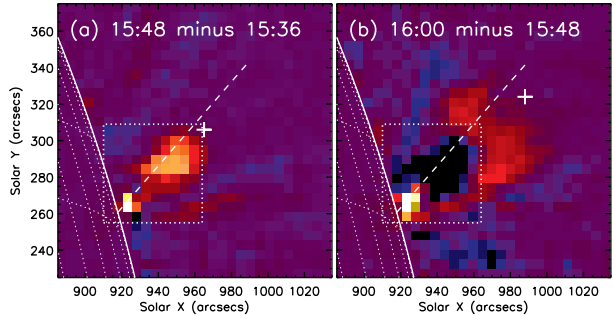}
 \caption[]{
 \soho EIT running difference images. The plus sign marks the leading edge of the eruption.
 The dashed line indicates the fiducial direction defined in Fig.~\ref{mapsII.eps}{\it c}.
 The white dotted boxes, identical to those in Figs.~\ref{mapsII.eps}{\it e}--\ref{mapsII.eps}{\it f}, 
 mark the FOV of Figs.~\ref{mapsII.eps}{\it a}--\ref{mapsII.eps}{\it d}.
 } \label{eit_rdiff.eps}
 \end{figure}

Phase~II consists of the impulsive (15:45--15:50~UT) and decay (15:50--16:14~UT) phases of the flare, 
during the former of which the prominence undergoes eruption. As mentioned above, 
the prominence is last seen	
by \trace at 15:44~UT (Fig.~\ref{traceMosaic.eps}{\it i}) 
just before the data gap. At the end of the gap (15:52~UT), a post-eruption 
{\it cusp} (Fig.~\ref{mapsII.eps}{\it d}) already appears at the original location of the prominence. 
Meanwhile, the prominence is last imaged in \Ha 
by ISOON (resolution: $1\arcsec$ vs.~\traceA's $0\farcs 5$, cadence: 5~minutes) at 15:40~UT, 
beyond which it is presumably too faint to be detected.		
Beyond the \trace and ISOON coverage, we located the leading edge of the eruption 
from \soho EIT 195 \AA\ running difference images (Fig.~\ref{eit_rdiff.eps}).
This edge is last detected at 16:00~UT at a projected height of $h=94\arcsec \pm 5\arcsec$ above the limb.
Differencing the heights at 15:48 and 16:00~UT gives a velocity of $29 \pm 7 \km \ps$, roughly
consistent with the final velocity of $27 \pm 4 \km \ps$ at 15:44~UT inferred from \trace data.
After another data gap (15:53-16:00~UT), we find in \trace 195~\AA\ images a series of post-flare loops
(with a cusp on the top) which appear to	
grow to larger loops at higher altitudes
(e.g., Figs.~\ref{mapsII.eps}{\it e} and \ref{mapsII.eps}{\it f}).
This is consistent with the classical flare model \citep[CSHKP,][]
{CarmichaelH1964psf..conf..451C, SturrockP1966Natur.211..695S, HirayamaT1974SoPh...34..323H, KoppR1976SoPh...50...85K}.
These observations indicate that the prominence erupts during the flare impulsive phase.
%
%
 \begin{figure*}[thbp]      
 \epsscale{0.38}  
 \plotone{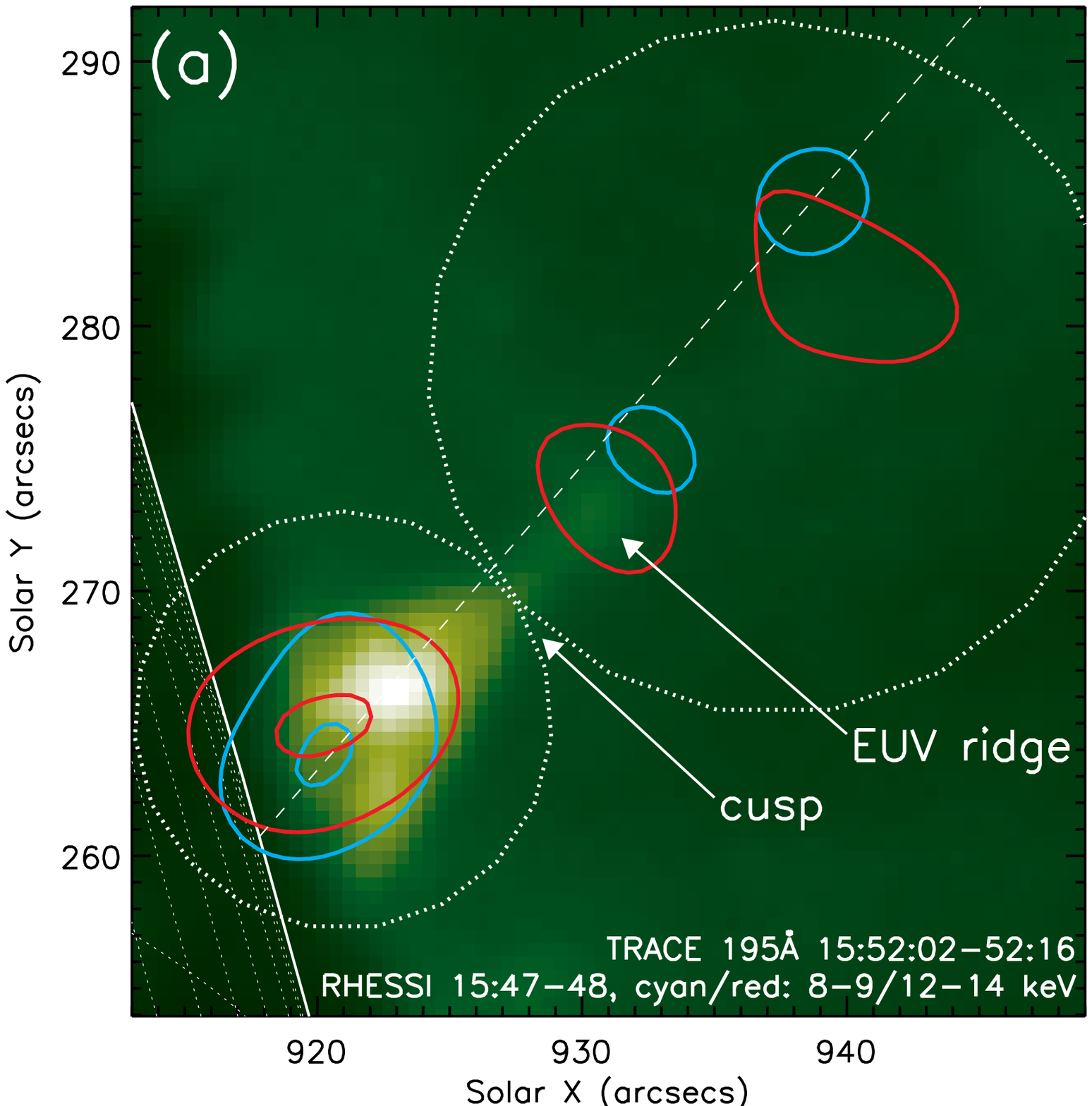}		
 \plotone{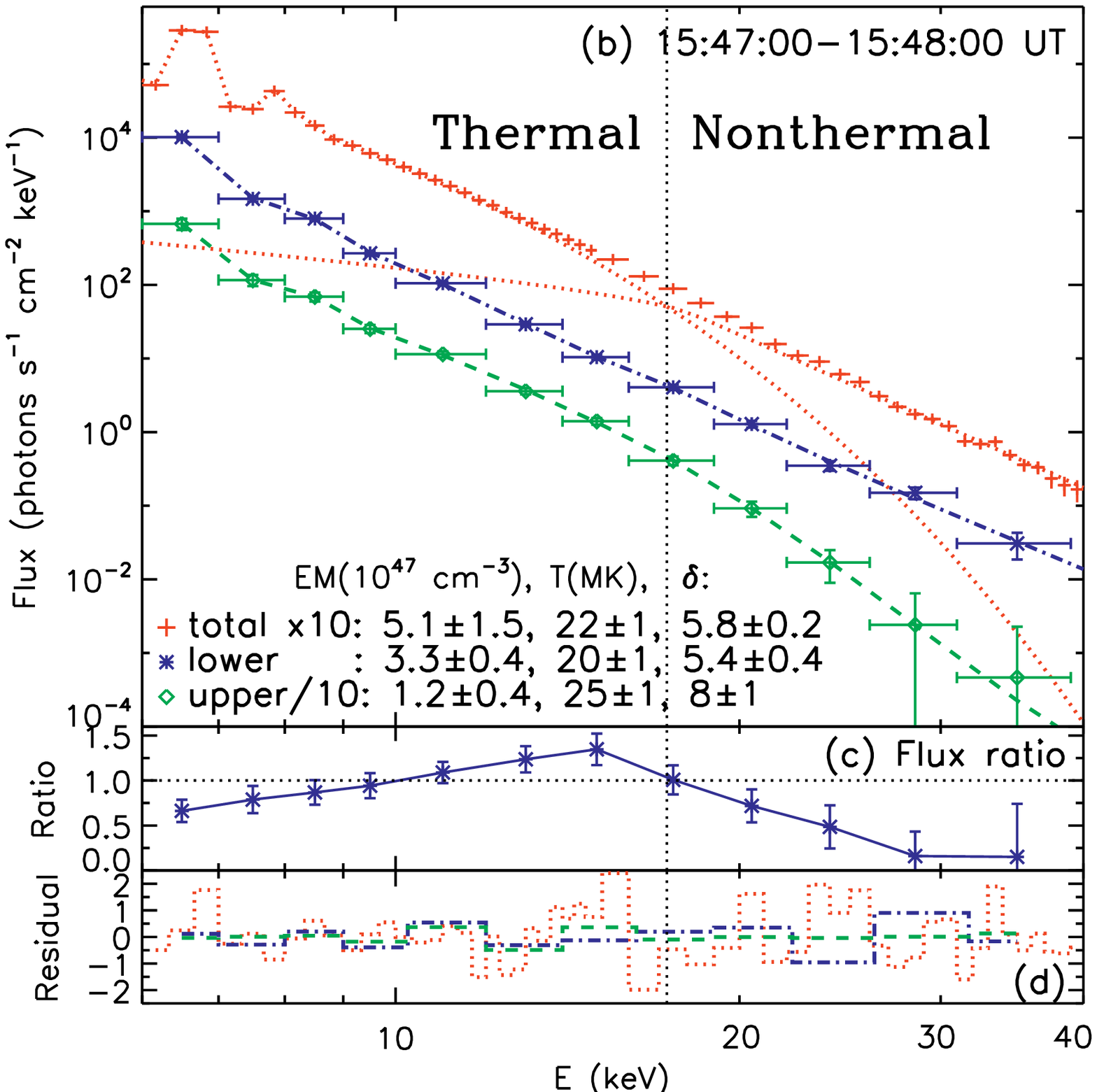}
 \plotone{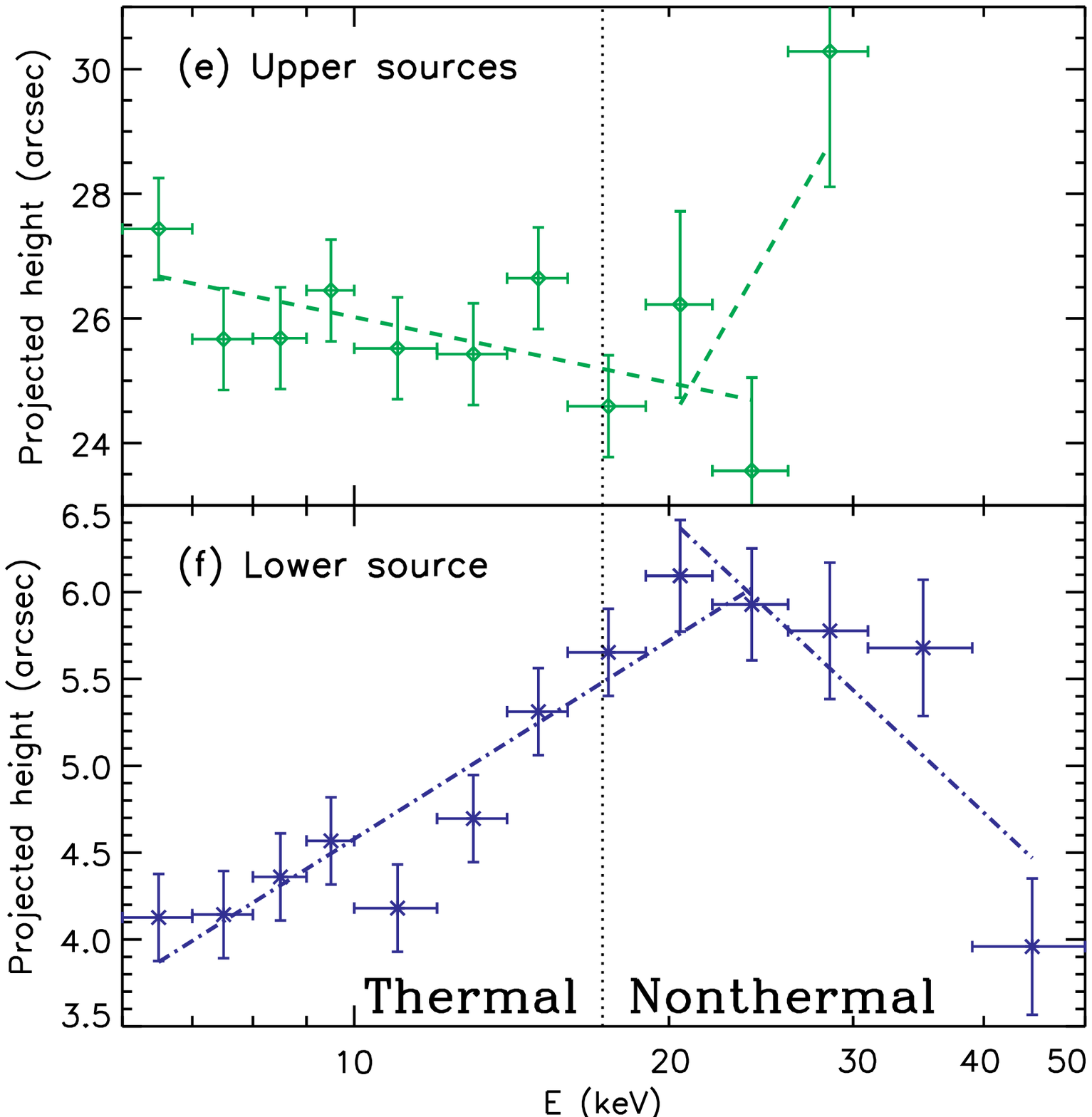}
 \caption[]{
 \hsi PIXON images and spectra at the peak (15:47:00--15:48:00~UT) of the impulsive phase.
   ({\it a}) Image contours at 8--9 ({\it cyan}) and 12--14~keV ({\it red}), 
 overplotted on a {\it later} \trace 195 \AA\ map at 15:52:02--15:52:16~UT as shown in Fig.~\ref{mapsII.eps}{\it d}.
 The contour levels are at 25\% and 90\% of the image maximum.
 The two white circles define the regions to obtain the spectra in the middle
 and the centroid positions on the right. 
 The dashed line is the fiducial direction defined in Fig.~\ref{mapsII.eps}{\it c}.
   ({\it b}) Spectra of the two regions (lower and upper) marked on the left and spatially integrated
 spectrum labeled as ``total". The ``total" and upper source spectra are shifted up and down
 by a factor of 10,	
 respectively.  The thermal and	
 nonthermal components of the ``total" spectrum are shown as dotted lines,	
 which cross each other at $E_{\rm cross}=17 \keV$ marked by the vertical dotted line here and on the right.
 These two components dominate below and above this critical energy.
 The total fits to the two spatially resolved spectra are shown as broken lines.
   ({\it c}) Flux ratio of the upper to lower sources.
   ({\it d}) Residuals of the three spectral fits normalized by 1-sigma uncertainties,
 in the same color and line styles as in panel {\it b}.
   ({\it e}) and ({\it f}) Projected heights of the centroids of the two regions defined on the left 
 as a function of energy. Broken lines are linear fits to the data.
 } \label{img-spec.eps}
 \end{figure*}

We now examine the spatial and spectral evolution of flare X-rays.	
Early (15:45:40--15:47:05~UT) during the impulsive phase after the thin attenuator moves in,	
the X-ray emission displays a complex morphology (e.g., Fig.~\ref{mapsII.eps}{\it a}) 
from which well-defined individual sources cannot be identified. 
For this period of time, we used the centroid of the overall X-ray emission 
to represent the source location (Fig.~\ref{vs-time.eps}{\it c}). At the peak 
(15:47:05--15:47:29~UT, Fig.~\ref{mapsII.eps}{\it b})
of the impulsive phase, two coronal sources are clearly imaged	
and their centroid positions are shown in Figure~\ref{vs-time.eps}{\it c}. 
Beyond this time, there are either multiple upper coronal sources 
(e.g., Figs.~\ref{mapsII.eps}{\it d}--\ref{mapsII.eps}{\it f}) or a single upper source much dimmer 
than 50\% of the image maximum, and thus we only show the centroid position of the lower source,
which gradually moves upward at a velocity of $7.4 \pm 0.2 \km \ps$ indicated by the linear fit.

\hsi spectral analysis indicates that the hot X-ray emitting plasma is rapidly
heated at the beginning of the impulsive phase and then undergoes gradual cooling.
The temperature decreases from $T=24 \pm 2 \MK$ at 15:45:40--15:46:40~UT
to $11.8 \pm 0.6 \MK$ at 16:10:36--16:12:36~UT (Fig.~\ref{vs-time.eps}{\it d}),
while the emission measure increases by about 10 times from $(2 \pm 1) \E{47}$ 
to $(1.9 \pm 0.9) \E{48} \pcmc$ at 15:48:40--15:49:40~UT and 
then remains roughly constant within a factor of 2  (Fig.~\ref{vs-time.eps}{\it e}). 
As expected, \goes data yield a lower temperature and higher emission measure, because 
\goes is more sensitive to cooler plasma than \hsiA.
In addition to the thermal emission, \hsi spectra indicate a	
nonthermal component during the impulsive phase.	
The spectral index $\delta$ of the power-law mean electron flux shows the usual soft-hard-soft evolution,
with the minimum (hardest) $\delta=5.8 \pm 0.2$ 	
occurring at the impulsive peak.

We also conducted imaging spectroscopic analysis for the interval of 15:47:00--15:48:00~UT
that covers the peak of the impulsive phase. To do this, we first made PIXON images in 13
progressively wider energy bins (from 1 to 11~keV) in the 6--50~keV range.
A sample of these images is shown in Figure~\ref{img-spec.eps}{\it a} as contours superimposed
on a {\it later} (15:52~UT) \trace 195 \AA\ image. There is a strong source located in the
low corona and two weak sources%
  \footnote{Note that photons from the two upper sources, primarily emitted during the second half
  of the integration time 15:47:00--15:48:00~UT, dominate over photons from the single upper source mostly emitted
  during the first half. Therefore the overall integration yields a structure of two upper sources, while
  the image shown in Fig.~\ref{mapsII.eps}{\it b} indicates only one upper source.
  }
higher up. As an independent confirmation, neighboring \goes SXI images show a teardrop-shaped
soft X-ray source, which is consistent with the convolution of the multiple \hsi sources
and SXI's gross resolution ($10 \arcsec$ FWHM, $5 \arcsec$ pixels). 
 We integrated the photon fluxes enclosed in the two 
white-dotted circles for the lower and upper coronal sources.  The resulting spectra
(see Fig.~\ref{img-spec.eps}{\it b}) were fitted with the same model 
(isothermal plus power-law mean electron flux) 	
as used for the spatially integrated spectra.
We find that the upper sources have a higher temperature ($T=25\pm 1$ vs.~$20 \pm 1 \MK$)	
but a lower emission measure than the lower source, and the temperature ($22\pm 1 \MK$) obtained from
the spatially integrated spectrum of the same interval sits in between.
On the other hand, the upper sources have a softer nonthermal component ($\delta=8 \pm 1$ vs.~$5.4 \pm 0.4$). 
These spectral differences can also be seen from the upper-to-lower flux ratio of the two spectra shown
in Figure~\ref{img-spec.eps}{\it c}. 

Close scrutiny of the images obtained above reveals an energy dependent source structure.
That is, the lower coronal source shifts to higher altitudes with increasing energies, while the
upper coronal sources shift to lower altitudes (e.g., Fig.~\ref{img-spec.eps}{\it a}).
At even higher energies this trend is reversed (see \S~\ref{sect_conclude} for a discussion).
To be quantitative, the projected heights of the centroids of the two regions 
are shown in Figures~\ref{img-spec.eps}{\it e} and \ref{img-spec.eps}{\it f} as a function of energy. 
We further applied linear fits to the data in two energy ranges, 6--26 and 19--50~keV,
in which the overlap of 19--26~keV is considered as the transition in between.	
The RMS deviations of the data from the fits are used to estimate the uncertainties.
The linear fits indicate that, from 6 to 26~keV, the projected heights of the lower centroid increases
by $2\farcs 2 \pm 0\farcs 4$, while that of the upper centroid decreases by $2 \farcs 0 \pm 1 \farcs 2$.
These observations are very similar to those of 
\citet[][see their Fig.~4]{LiuW_2LT.2008ApJ...676..704L} and of \citet{SuiL2003ApJ...596L.251S},
and suggest that magnetic reconnection occurs between the two regions, 
presumably in a current sheet which is oriented along the line connecting the two centroids. 	
This scenario is further supported by our new observation of the cusp and the {\it current-sheet-like}
feature or {\it ridge}%
  \footnote{	
  \citet{Sui.enigma.2006ApJ...646..605S} found a coronal X-ray source at the location 
  where an EUV ridge (above a cusp) appeared later. That source, however, 
  was too weak to allow detailed imaging or spectral analysis.}
above it seen at 195 \AA\ by \trace 4 minutes later%
 \footnote{There is no simultaneous \trace image during the impulsive phase due to the data gap noted earlier.}
(Fig.~\ref{img-spec.eps}{\it a}).
(The width of the ridge is $\sim$$3\arcsec$ or 2200~km.)
In addition, the three \hsi X-ray sources are nearly in a straight line along the EUV ridge.

Nearly simultaneous \hsi and \trace images (see Fig.~\ref{mapsII.eps}{\it d}) taken around 15:52~UT,
2~minutes after the impulsive phase, show a similar situation except that all \hsi sources 
have shifted to higher altitudes and the upper sources have become even weaker. 
We find that the centroid of the lower coronal source at 6--10~keV is located 
$7\arcsec \pm 1 \arcsec$ below the cusp and cospatial with the EUV brightening, while the upper sources 
are above the cusp and dispersed along	
the extended EUV ridge.
(Earlier at the impulsive peak during the \trace data gap, we speculate that an EUV cusp and ridge also exist
at similar positions relative to the X-ray sources, but at lower altitudes.)
According to the spatially integrated spectrum at this time, 
the X-ray emitting plasma has a temperature of $T=17.9 \pm 0.4 \MK$.
The cospatial 195~\AA\ emission can be explained by the high-temperature response of this \trace
channel due to Fe~XXIV line (peak at 16~MK) emission \citep{PhillipsKen.highT195.2005ApJ...626.1110P}.
Later in the decay phase (Figs.~\ref{mapsII.eps}{\it e} and \ref{mapsII.eps}{\it f}), 
the lower (hot) X-ray source is located above the cooler EUV post-flare loops as seen 
in many other flares \citep[e.g.,][]{GallagherP2002SoPh..210..341G}. There are also multiple X-ray sources
(Figs.~\ref{mapsII.eps}{\it e} and \ref{mapsII.eps}{\it f})      
at significantly higher altitudes that are located away from the direction of the EUV current-sheet-like feature
and from the fiducial direction of motion.
We speculate that these sources are at the tops of hot post-flare loops. They are heated
either {\it in-situ} as a result of local magnetic reconnection induced by the prominence eruption,
or from chromospheric evaporation in response to electron beam or conductive heating at the footpoints.

\section{Summary and Discussion}	
\label{sect_conclude}

We have presented \hsi X-ray and \trace EUV observations of a C7.7 flare accompanying a prominence eruption. 
This event exhibits two distinct phases whose characteristics are summarized as follows:


{\bf 1.}~Phase~I is marked with the preheating of the flare and the activation and rise of the prominence. 
Phase~II includes the flare impulsive phase, which accompanies the prominence eruption, 
and the flare decay phase (Fig.~\ref{vs-time.eps}{\it a}). Thermal X-rays are present in both Phases~I and II with
peak temperatures of $T=23\pm 1$	
and  $24 \pm 2 \MK$, respectively. There is		
nonthermal emission during the impulsive phase with the minimum (hardest) power-law index 
of the mean electron flux $\delta=5.8 \pm 0.2$ (Fig.~\ref{vs-time.eps}{\it d}).

{\bf 2.}~During Phase~I there are three episodes of morphological changes of the X-ray emission from
a single source to a double source (Figs.~\ref{vs-time.eps}{\it c} and \ref{traceMosaic.eps}).
Each episode coincides with an increased X-ray flux and plasma temperature 
(Figs.~\ref{vs-time.eps}{\it b} and \ref{vs-time.eps}{\it d}),
suggestive of faster magnetic reconnection or a larger energy release rate. However, the prominence motion
does not show an obvious corresponding episodic behavior, except that its eruption
occurs during the flare impulsive phase.

{\bf 3.}~Most of the time, the X-ray emission is located primarily {\it below} the apex of the prominence.
Only for $\sim$4 minutes (15:33--15:37~UT) during the middle stage of Phase~I, some or all X-rays 
are emitted from above the apex (Fig.~\ref{vs-time.eps}{\it c}). 
In particular, the initial X-ray emission occurs at the northern leg of the prominence	
(Fig.~\ref{traceMosaic.eps}{\it a}).		
Since X-rays at this time are primarily thermal emission, a signature of hot plasma presumably
heated as a consequence of magnetic reconnection, these observations
support the tether-cutting model \citep{MooreRoumeliotis.origTether-cutting.1992LNP...399...69M} 
over the breakout model \citep{AntiochosS.breakout.1998ApJ...502L.181A} for triggering solar eruptions.

{\bf 4.}~During the impulsive phase, the X-ray sources display an energy-dependent structure 
(Figs.~\ref{img-spec.eps}{\it e} and \ref{img-spec.eps}{\it f}). That is, higher
energy emission is closer together toward the EUV cusp in the thermal regime ($\lesssim$20~keV),
with a reversed trend in the nonthermal regime (see the discussion below).
In addition, there is a bright EUV ridge	
extending from the cusp to higher altitudes,
which is aligned with the multiple X-ray sources (Fig.~\ref{img-spec.eps}{\it a}).
These observations suggest the existence of a current sheet, most likely trailing the erupting
prominence, as predicted by the classical CSHKP flare model.
Compared with previously reported X-ray signatures of current sheets \citep[e.g.,][]{SuiL2003ApJ...596L.251S,
Sui.enigma.2006ApJ...646..605S, LiuW_2LT.2008ApJ...676..704L}, the combination of these X-ray and EUV observations
represents another example with additional information.		

{\bf 5.}~During the impulsive phase, the upper coronal sources have a higher temperature ($T=25 \pm 1$ vs.~$20 \pm 1 \MK$) 
but lower emission measure (${\rm EM}=[1.2 \pm 0.4]$ vs.~$[3.3 \pm 0.4] \E{47} \pcmc $)
than the lower coronal source (Fig.~\ref{img-spec.eps}{\it b}). This is expected and consistent with the finding of 
\citet{LiuW_2LT.2008ApJ...676..704L}, who proposed differences in magnetic connectivity to be the primary cause. 
On the other hand, the nonthermal spectra of \citet{LiuW_2LT.2008ApJ...676..704L} show similar power-law indexes in
the two regions. However, 	
the upper sources in our case show much softer ($\delta=8 \pm 1$ vs.~$5.4 \pm 0.4$) nonthermal emission,
suggestive of a particle acceleration rate or efficiency different from that of the lower source.
An alternative explanation is that the lower source is thick-target emission, while the upper sources
are thin-target. This is because the difference ($\Delta \delta$) of the electron power-law indexes
of the two regions equals that ($\Delta \gamma$) of the photon indexes,
since the functional forms of the fits are identical for both regions.
It then follows $\Delta \gamma=\Delta \delta=2.6 \pm 1.1$, which is consistent 	
with the predicted value of $\Delta \gamma=2$ between the thin- and thick-target cases 
\citep[e.g.,][]{BrownJ1971SoPh, LinR.HudsonH.HXR1971SoPh...17..412L, PetrosianV1973ApJ...186..291P}.
This interpretation qualitatively agrees
with the inferred higher emission measure and possibly higher density for the lower source.



In what follows, we further discuss the implications of some of our observations:

{\bf 1.}~The {\bf elongation and splitting} of a single X-ray source into a double source was previously observed
near the onset of the impulsive phase \citep{SuiL2003ApJ...596L.251S, VeronigA2006A&A...446..675V, WangT.SuiL2007ApJ,
LiuW_2LT.2008ApJ...676..704L}, similar to what happens near 15:47~UT in this flare. 
This was interpreted as the signature of the formation of a large-scale current sheet through
the transformation from an X null point to a double-Y shaped magnetic topology \citep{SuiL2003ApJ...596L.251S},
or as evidence of the development of fast reconnection occurring in a pre-existing current sheet \citep{LiuW_2LT.2008ApJ...676..704L}. 
It is likely that, in this event, the three episodes of source splitting during the flare preheating 
and prominence activation phase reflect the same signature of reconnection in a current sheet.
However, in the previously reported events, such splitting occurs around the onset of the impulsive phase
and occurs only once, while in this event it happens prior to the impulsive phase and it happens three times.
Possible reasons behind this difference are as follows.

In this and the previously reported larger (M--X class) events, the source splitting at the beginning of the impulsive
phase takes place in a simple, {\it large-scale} current sheet at the onset of fast reconnection as predicted in the 
classical CSHKP picture. Once it happens, reconnection goes into a runaway situation, leading to {\it explosive}
energy release. This scenario is supported by the observations (see Fig.~\ref{vs-time.eps}{\it c}) of this event that
 (1) the impulsive phase does not start until the erupting prominence is significantly high 
($\gtrsim 20\arcsec$ or 15~Mm) above the limb and thus presumably until a large-scale current sheet 	
is created in its wake;	
 and (2) early in the impulsive phase the X-ray emission (presumably near the reconnection site) is located
below the prominence apex and thus possibly situated in the 	
large-scale current sheet.

In contrast, it is likely that the source splitting prior to the impulsive phase in this event takes place
episodically in {\it localized}, {\it small-scale} current sheets located in the prominence or its vicinity. 
Each episode occurs in a different current sheet, which is shortly dissipated away by reconnection, 
giving rise to {\it moderate} energy release, an increased plasma temperature, and an enhanced X-ray flux 
during the preheating phase.
Such episodic reconnection could happen as: 
 (1) tether-cutting reconnection in small-scale current sheets between oppositely aligned field lines in a sheared core 
\citep{MooreRoumeliotis.origTether-cutting.1992LNP...399...69M},
 or (2) reconnection in curved current sheets around a kinked, pre-existing twisted flux rope
\citep{Kliem_curr-sheet-sigmoid-kink.2004A&A...413L..23K, FanY.kink.2005ApJ...630..543F}.
  In the former case, a flux rope could be formed via reconnection 
\citep{vanBallegooijenMartens.promin.1989ApJ...343..971V}, which, at the same time, 
can increase the twist or helicity of the rope. The helical structure seen in the late stage of the filament 
activation (e.g., Fig.~\ref{traceMosaic.eps}{\it g}) is a possible indicator of the increased twist.
  In the latter case, the kink instability converts the twist helicity into the writhe helicity of
the flux rope. The observed helical feature may be the manifestation of the writhe.
(3) There is a third possibility that the above two scenarios are both present, but in different stages
of the evolution. That is, early in the preheating phase, tether-cutting reconnection
could episodically develop and increase the twist of the flux rope. 
Once the twist exceeds a threshold, the flux rope becomes kink unstable,
leading to the transition to the impulsive phase. 	
Our observations do not allow us to differentiating between these three possibilities,
partly because of loss of information of activity in the lower atmosphere behind the limb.
  After the initiation, however, the interlaced periods of X-ray emission
being below and above the prominence apex (Fig.~\ref{vs-time.eps}{\it c}) suggest a circular chain of
positive feedback from both reconnection and prominence activity \citep{MooreSterling.2006GMS...165...43M, 
ChiforC.filamentErupt.2006A&A...458..965C}.

{\bf 2.}~The {\bf reversal} of the trend in the energy dependence of the centroid heights 	
(Fig.~\ref{img-spec.eps}{\it c}) confirms the result discovered by \citet{LiuW_2LT.2008ApJ...676..704L}. 
According to the spatially integrated spectrum shown in Figure~\ref{img-spec.eps}{\it b},
the thermal and nonthermal components intersect at $E_{\rm cross}=17 \pm 1 \keV$.
(1) Below this energy, thermal emission dominates and thus the energy-dependent source structure
simply implies X-ray emitting plasma of higher temperatures (but lower emission measures) being closer
to the reconnection site located between the lower and upper sources \citep{SuiL2003ApJ...596L.251S}.
This was interpreted and generalized by \citet{LiuW_2LT.2008ApJ...676..704L} in the framework of stochastic acceleration
by turbulence or plasma waves \citep{PetrosianV2004ApJ...610..550P}.
Such a scenario for the lower coronal (loop-top) source alone was also modeled with a collapsing trap
\citep{Karlicky.Kosugi.collps-trap.2004A&A...419.1159K} by \citet{VeronigA2006A&A...446..675V}.
(2) Above $E_{\rm cross}$, the emission is mainly nonthermal and the reversal of the trend at $\sim$20~keV
indicates a transition from thermal to nonthermal dominance. It may further imply
increasing importance of transport effects of {\it nonthermal} electrons \citep{LiuW_2LT.2008ApJ...676..704L}.
That is, higher energy electrons have larger stopping column densities and they tend to produce 
bremsstrahlung at greater distances from their acceleration site.
This is more likely to be true for the lower source, given its stronger signature of the trend reversal
(Fig.~\ref{img-spec.eps}{\it f}) and the possibility of thick-target emission noted above (Item~5).
\citet{SuiL2003ApJ...596L.251S} also found a jump of the centroid position
at $\sim$17~keV and interpreted it as a thermal-to-nonthermal transition.	

We are currently analyzing data from the \soho Solar Ultraviolet Measurements of Emitted
Radiation (SUMER) for this and other events from this active region.
The inferred temperatures and Doppler velocities may shed light on the nature of 
the high coronal X-ray sources mentioned
earlier (see Figs.~\ref{mapsII.eps}{\it e} and \ref{mapsII.eps}{\it f}), and will
be presented in a separate publication. The observations presented here also point to a promising direction
for future investigations, that is, to closely follow the evolution of the morphology and spectra
of the X-ray emission and of the accompanied prominence activity. We look forward to carrying out
such studies using better data from current and future missions such as the
{\it Solar Dynamics Observatory} ({\it SDO}).


\acknowledgements
{W.~Liu was supported by an appointment to the NASA Postdoctoral Program at Goddard Space Flight Center, 
administered by Oak Ridge Associated Universities through a contract with NASA. 
T. Wang's work was supported by NRL grant N00173-06-1-G033 and NASA grant NNX08AE44G.
B. Dennis and G. Holman were supported in part by the NASA Heliophysics Guest Investigator Program.
We thank Holly Gilbert, Ken Phillips, Jing Li, David Webb, Nariaki Nitta, and Jiong Qiu 
for fruitful discussions and/or various help. 
We also thank the anonymous referee for constructive comments.
The Improved Solar Observing Optical Network (ISOON) project is a collaboration between the 
Air Force Research Laboratory Space Vehicles Directorate and the National Solar Observatory.
}


{\scriptsize
\bibliography{bib/ads_all_edit,bib/LiuW-group,bib/Liu-Wei}

\begin{thebibliography}{47}
\expandafter\ifx\csname natexlab\endcsname\relax\def\natexlab#1{#1}\fi

\bibitem[{{Alexander} {et~al.}(2006){Alexander}, {Liu}, \&
  {Gilbert}}]{AlexanderLiuGilbert.failed.2006ApJ...653..719A}
{Alexander}, D., {Liu}, R., \& {Gilbert}, H.~R. 2006, \apj, 653, 719

\bibitem[{{Antiochos}(1998)}]{AntiochosS.breakout.1998ApJ...502L.181A}
{Antiochos}, S.~K. 1998, \apjl, 502, L181

\bibitem[{{Brown}(1971)}]{BrownJ1971SoPh}
{Brown}, J.~C. 1971, \solphys, 18, 489

\bibitem[{{Brown} {et~al.}(2003){Brown}, {Emslie}, \&
  {Kontar}}]{Brown.meanElecflux.2003ApJ...595L.115B}
{Brown}, J.~C., {Emslie}, A.~G., \& {Kontar}, E.~P. 2003, \apjl, 595, L115

\bibitem[{{Carmichael}(1964)}]{CarmichaelH1964psf..conf..451C}
{Carmichael}, H. 1964, in The Physics of Solar Flares, ed. W.~N. {Hess}, 451

\bibitem[{{Chifor} {et~al.}(2006){Chifor}, {Mason}, {Tripathi}, {Isobe}, \&
  {Asai}}]{ChiforC.filamentErupt.2006A&A...458..965C}
{Chifor}, C., {Mason}, H.~E., {Tripathi}, D., {Isobe}, H., \& {Asai}, A. 2006,
  \aap, 458, 965

\bibitem[{{Chifor} {et~al.}(2007){Chifor}, {Tripathi}, {Mason}, \&
  {Dennis}}]{ChiforC.XRprecursor.2007A&A...472..967C}
{Chifor}, C., {Tripathi}, D., {Mason}, H.~E., \& {Dennis}, B.~R. 2007, \aap,
  472, 967

\bibitem[{{Fan}(2005)}]{FanY.kink.2005ApJ...630..543F}
{Fan}, Y. 2005, \apj, 630, 543

\bibitem[{{Fan} \& {Gibson}(2003)}]{FanGibson.emrgKink.2003ApJ...589L.105F}
{Fan}, Y. \& {Gibson}, S.~E. 2003, \apjl, 589, L105

\bibitem[{{Gallagher} {et~al.}(2002){Gallagher}, {Dennis}, {Krucker},
  {Schwartz}, \& {Tolbert}}]{GallagherP2002SoPh..210..341G}
{Gallagher}, P.~T., {Dennis}, B.~R., {Krucker}, S., {Schwartz}, R.~A., \&
  {Tolbert}, A.~K. 2002, \solphys, 210, 341

\bibitem[{{Handy} {et~al.}(1999){Handy}, {Tarbell}, {Wolfson}, {Korendyke}, \&
  {Vourlidas}}]{HandyB.Lyman-alpha.1999SoPh..190..351H}
{Handy}, B.~N., {Tarbell}, T.~D., {Wolfson}, C.~J., {Korendyke}, C.~M., \&
  {Vourlidas}, A. 1999, \solphys, 190, 351

\bibitem[{{Harrison}(1986)}]{Harrison.precursor.1986A&A...162..283H}
{Harrison}, R.~A. 1986, \aap, 162, 283

\bibitem[{{Harrison} {et~al.}(1985){Harrison}, {Waggett}, {Bentley},
  {Phillips}, {Bruner}, {Dryer}, \&
  {Simnett}}]{Harrison.etal.precursor.1985SoPh...97..387H}
{Harrison}, R.~A., {Waggett}, P.~W., {Bentley}, R.~D., {Phillips}, K.~J.~H.,
  {Bruner}, M., {Dryer}, M., \& {Simnett}, G.~M. 1985, \solphys, 97, 387

\bibitem[{{Hirayama}(1974)}]{HirayamaT1974SoPh...34..323H}
{Hirayama}, T. 1974, \solphys, 34, 323

\bibitem[{{Hurford} {et~al.}(2002){Hurford}, {Schmahl}, {Schwartz}, {Conway},
  {Aschwanden}, {Csillaghy}, {Dennis}, {Johns-Krull},
  {et~al.}}]{HurfordG2002SoPh..210...61H}
{Hurford}, G.~J., {Schmahl}, E.~J., {Schwartz}, R.~A., {Conway}, A.~J.,
  {Aschwanden}, M.~J., {Csillaghy}, A., {Dennis}, B.~R., {Johns-Krull}, C.,
  {et~al.} 2002, \solphys, 210, 61

\bibitem[{{Ji} {et~al.}(2003){Ji}, {Wang}, {Schmahl}, {Moon}, \&
  {Jiang}}]{JiH.FailedEruption.2003ApJ...595L.135J}
{Ji}, H., {Wang}, H., {Schmahl}, E.~J., {Moon}, Y.-J., \& {Jiang}, Y. 2003,
  \apjl, 595, L135

\bibitem[{{Karlick{\'y}} \&
  {Kosugi}(2004)}]{Karlicky.Kosugi.collps-trap.2004A&A...419.1159K}
{Karlick{\'y}}, M. \& {Kosugi}, T. 2004, \aap, 419, 1159

\bibitem[{{Kliem} {et~al.}(2004){Kliem}, {Titov}, \&
  {T{\"o}r{\"o}k}}]{Kliem_curr-sheet-sigmoid-kink.2004A&A...413L..23K}
{Kliem}, B., {Titov}, V.~S., \& {T{\"o}r{\"o}k}, T. 2004, \aap, 413, L23

\bibitem[{{Kopp} \& {Pneuman}(1976)}]{KoppR1976SoPh...50...85K}
{Kopp}, R.~A. \& {Pneuman}, G.~W. 1976, \solphys, 50, 85

\bibitem[{{Li} {et~al.}(2005){Li}, {Mickey}, \&
  {LaBonte}}]{LiJing.20715X3.2005ApJ...620.1092L}
{Li}, J., {Mickey}, D.~L., \& {LaBonte}, B.~J. 2005, \apj, 620, 1092

\bibitem[{{Lin} \& {Hudson}(1971)}]{LinR.HudsonH.HXR1971SoPh...17..412L}
{Lin}, R.~P. \& {Hudson}, H.~S. 1971, \solphys, 17, 412

\bibitem[{{Liu} \& {Alexander}(2009)}]{LiuAlexander.kink.HXR.2009ApJ}
{Liu}, R. \& {Alexander}, D. 2009, accepted by \apj

\bibitem[{{Liu} {et~al.}(2009){Liu}, {Petrosian}, {Dennis}, \&
  {Holman}}]{LiuW_FPAsym_2009ApJ...693..847L}
{Liu}, W., {Petrosian}, V., {Dennis}, B.~R., \& {Holman}, G.~D. 2009, \apj,
  693, 847

\bibitem[{{Liu} {et~al.}(2008){Liu}, {Petrosian}, {Dennis}, \&
  {Jiang}}]{LiuW_2LT.2008ApJ...676..704L}
{Liu}, W., {Petrosian}, V., {Dennis}, B.~R., \& {Jiang}, Y.~W. 2008, \apj, 676,
  704

\bibitem[{{Liu} \& {Kurokawa}(2004)}]{LiuKurokawa.filament.2004PASJ...56..497L}
{Liu}, Y. \& {Kurokawa}, H. 2004, \pasj, 56, 497

\bibitem[{{Low} \& {Zhang}(2002)}]{LowZhang.2classCME.2002ApJ...564L..53L}
{Low}, B.~C. \& {Zhang}, M. 2002, \apjl, 564, L53

\bibitem[{{Metcalf} {et~al.}(1996){Metcalf}, {Hudson}, {Kosugi}, {Puetter}, \&
  {Pina}}]{MetcalfT1996ApJ...466..585M}
{Metcalf}, T.~R., {Hudson}, H.~S., {Kosugi}, T., {Puetter}, R.~C., \& {Pina},
  R.~K. 1996, \apj, 466, 585

\bibitem[{{Moore} \&
  {Roumeliotis}(1992)}]{MooreRoumeliotis.origTether-cutting.1992LNP...399...69%
M}
{Moore}, R.~L. \& {Roumeliotis}, G. 1992, in Lecture Notes in Physics, Berlin
  Springer Verlag, Vol. 399, IAU Colloq. 133: Eruptive Solar Flares, ed.
  Z.~{Svestka}, B.~V. {Jackson}, \& M.~E. {Machado}, 69

\bibitem[{{Moore} \& {Sterling}(2006)}]{MooreSterling.2006GMS...165...43M}
{Moore}, R.~L. \& {Sterling}, A.~C. 2006, AGU Geophys.~Monog.~Series: Solar
  Eruptions and Energetic Particles, Eds. N. Gopalswamy, R. Mewaldt, \& J.
  Torsti (Washington DC), 165, 43

\bibitem[{{Moore} {et~al.}(2001){Moore}, {Sterling}, {Hudson}, \&
  {Lemen}}]{MooreR.tether-cutting.2001ApJ...552..833M}
{Moore}, R.~L., {Sterling}, A.~C., {Hudson}, H.~S., \& {Lemen}, J.~R. 2001,
  \apj, 552, 833

\bibitem[{{Petrosian}(1973)}]{PetrosianV1973ApJ...186..291P}
{Petrosian}, V. 1973, \apj, 186, 291

\bibitem[{{Petrosian} \& {Liu}(2004)}]{PetrosianV2004ApJ...610..550P}
{Petrosian}, V. \& {Liu}, S. 2004, \apj, 610, 550

\bibitem[{{Phillips} {et~al.}(2005){Phillips}, {Chifor}, \&
  {Landi}}]{PhillipsKen.highT195.2005ApJ...626.1110P}
{Phillips}, K.~J.~H., {Chifor}, C., \& {Landi}, E. 2005, \apj, 626, 1110

\bibitem[{{Smith} {et~al.}(2002){Smith}, {Lin}, {Turin}, {Curtis}, {Primbsch},
  {Campbell}, {Abiad}, {Schroeder}, {et~al.}}]{SmithD2002SoPh..210...33S}
{Smith}, D.~M., {Lin}, R.~P., {Turin}, P., {Curtis}, D.~W., {Primbsch}, J.~H.,
  {Campbell}, R.~D., {Abiad}, R., {Schroeder}, P., {et~al.} 2002, \solphys,
  210, 33

\bibitem[{{Sterling} \&
  {Moore}(2004{\natexlab{a}})}]{SterlingMoore.external-recon.2004ApJ...602.102%
4S}
{Sterling}, A.~C. \& {Moore}, R.~L. 2004{\natexlab{a}}, \apj, 602, 1024

\bibitem[{{Sterling} \&
  {Moore}(2004{\natexlab{b}})}]{SterlingMoore.exter-internal-recon.2004ApJ...6%
13.1221S}
---. 2004{\natexlab{b}}, \apj, 613, 1221

\bibitem[{{Sturrock}(1966)}]{SturrockP1966Natur.211..695S}
{Sturrock}, P.~A. 1966, \nat, 211, 695

\bibitem[{{Sui} \& {Holman}(2003)}]{SuiL2003ApJ...596L.251S}
{Sui}, L. \& {Holman}, G.~D. 2003, \apjl, 596, L251

\bibitem[{{Sui} {et~al.}(2006){Sui}, {Holman}, \&
  {Dennis}}]{Sui.enigma.2006ApJ...646..605S}
{Sui}, L., {Holman}, G.~D., \& {Dennis}, B.~R. 2006, \apj, 646, 605

\bibitem[{{Tandberg-Hanssen}(1995)}]{Tandberg-HanssenE.prominenceBook.1995nsp.%
.book.....T}
{Tandberg-Hanssen}, E. 1995, {The nature of solar prominences} (Dordrecht;
  Boston: Kluwer)

\bibitem[{{T{\"o}r{\"o}k} \&
  {Kliem}(2005)}]{Torok.Kliem.kink.2005ApJ...630L..97T}
{T{\"o}r{\"o}k}, T. \& {Kliem}, B. 2005, \apjl, 630, L97

\bibitem[{{van Ballegooijen} \&
  {Martens}(1989)}]{vanBallegooijenMartens.promin.1989ApJ...343..971V}
{van Ballegooijen}, A.~A. \& {Martens}, P.~C.~H. 1989, \apj, 343, 971

\bibitem[{{van Hoven} \&
  {Hurford}(1984)}]{vanHoven.Hurford.precursor.1984AdSpR...4...95V}
{van Hoven}, G. \& {Hurford}, G.~J. 1984, Advances in Space Research, 4, 95

\bibitem[{{Veronig} {et~al.}(2006){Veronig}, {Karlick{\'y}}, {Vr{\v s}nak},
  {Temmer}, {Magdaleni{\'c}}, {Dennis}, {Otruba}, \&
  {P{\"o}tzi}}]{VeronigA2006A&A...446..675V}
{Veronig}, A.~M., {Karlick{\'y}}, M., {Vr{\v s}nak}, B., {Temmer}, M.,
  {Magdaleni{\'c}}, J., {Dennis}, B.~R., {Otruba}, W., \& {P{\"o}tzi}, W. 2006,
  \aap, 446, 675

\bibitem[{{Wang} {et~al.}(2007){Wang}, {Sui}, \& {Qiu}}]{WangT.SuiL2007ApJ}
{Wang}, T.~J., {Sui}, L., \& {Qiu}, J. 2007, \apjl, 661, L207

\bibitem[{{Wang} {et~al.}(2002){Wang}, {Yan}, {Wang}, {Kurokawa}, \&
  {Shibata}}]{WangTJ.baldpatch.2002ApJ...572..580W}
{Wang}, T.~J., {Yan}, Y., {Wang}, J., {Kurokawa}, H., \& {Shibata}, K. 2002,
  \apj, 572, 580

\bibitem[{{Williams} {et~al.}(2005){Williams}, {T{\"o}r{\"o}k}, {D{\'e}moulin},
  {van Driel-Gesztelyi}, \&
  {Kliem}}]{WilliamsDR.kink-promin.2005ApJ...628L.163W}
{Williams}, D.~R., {T{\"o}r{\"o}k}, T., {D{\'e}moulin}, P., {van
  Driel-Gesztelyi}, L., \& {Kliem}, B. 2005, \apjl, 628, L163

\end{thebibliography}
}




\end{document}